\documentclass[bimj,fleqn]{w-art}

\usepackage{times}
\usepackage{w-thm}
\usepackage{texab}
\usepackage{lscape}
\usepackage{graphicx}
\usepackage{multirow}
\usepackage{url}  % Formatting web addresses  
\usepackage{multicol}   %Columns
\usepackage[authoryear]{natbib}
\usepackage{algorithm2e}
\usepackage{graphicx}
\begin{document}
%% The information for the title page will be placed between
%%    \begin{document} and \maketitle. The order of most entries
%%    is determined by the class file and can not be changed by
%%    rearranging them. The maketitle command follows after the
%%    abstract.
%%
%%    Most of the following commands will be completed by the publisher.
%%
%%    The copyrightyear is defined in the .clo file as the first argument
%%    of the copyrightinfo command. If the copyrightyear differs from that
%%    value it might be adjusted by the following definition:
%%
%% \renewcommand{\copyrightyear}{2004}% uncomment to change the copyrightyear.
%%
\DOIsuffix{bimj.DOIsuffix}
%%
%% issueinfo for the header line
\Volume{46}
\Issue{1}
\Year{2004}
%%
%%    First and last pagenumber of the article. If the option
%%    'autolastpage' is set (default) the second argument may be left empty.
\pagespan{1}{}
%%
%%    Dates will be filled in by the publisher. The 'reviseddate' and
%%    'dateposted' (Published online) entry may be left empty.
\Receiveddate{2 April 2009}
\Reviseddate{}
\Accepteddate{}
\Dateposted{}
\keywords{Categorical Data, Graphical Model, Group Lasso, Log-Linear Models, Sparse Contingency Tables.}

%% \pretitle{Editor's Choice}

%% We have a short and a long form for the title. The short form
%% (optional argument) goes into the running head.

\title[Decomposition and Model Selection for Large Contingency Tables]{Decomposition and Model Selection for Large Contingency Tables}

%% Please do not enter footnotes or \inst{}-notes into the optional
%% argument of the author command. The optional argument will go into
%% the header.  If there is only one address the marker \inst{x} may be
%% omitted.

%% Information for the first author.
\author[C. Dahinden]{Corinne Dahinden\inst{1,2}} \address[\inst{1}]{Seminar f\"ur Statistik, ETH Z\"urich, CH-8092 Z\"urich, Switzerland}
%%
%%    Information for the second author
\author[M. Kalisch]{Markus Kalisch\footnote{Corresponding
     author: e-mail: {\sf kalisch@stat.math.ethz.ch}, Phone: +41\,44\,632\,3435,
     Fax: +41\,44\,632\,1228}\inst{1}}
\author[P. B\"uhlmann]{Peter B\"uhlmann\inst{1,2}}
\address[\inst{2}]{Competence Center for Systems Physiology and Metabolic
    Diseases, ETH Z\"urich, CH-8093 Z\"urich, Switzerland}
%%
%%    Information for the third author
%%
%%    \dedicatory{This is a dedicatory.}
\begin{abstract}
  Large contingency tables summarizing categorical variables arise in many
  areas. One example is in biology, where large numbers of biomarkers are
  cross-tabulated according to their discrete expression level.
  Interactions of the variables are of great interest and are generally
  studied with log-linear models. The structure of a log-linear model can
  be visually represented by a graph from which the conditional
  independence structure can then be easily read off.  However, since the
  number of parameters in a saturated model grows exponentially in the
  number of variables, this generally comes with a heavy computational
  burden. Even if we restrict ourselves to models of lower order
  interactions or other sparse structures we are faced with the problem of
  a large number of cells which plays the role of sample size. This is in
  sharp contrast to high-dimensional regression or classification
  procedures because, in addition to a high-dimensional parameter,
  we also have to deal with the analogue of a huge sample
  size. Furthermore, high-dimensional tables naturally feature a large
  number of sampling zeros which often leads to the nonexistence of the
  maximum likelihood estimate. We therefore present a decomposition
  approach, where we first divide the problem into several
  lower-dimensional problems and then combine these to form a global
  solution. Our methodology is computationally feasible for log-linear
  interaction models with many categorical variables each or some of them
  having many levels. We demonstrate the proposed method on simulated
  data and apply it to a bio-medical problem in cancer
  research.\end{abstract}
%% maketitle must follow the abstract.
\maketitle                   % Produces the title.

%% If there is not enough space inside the running head
%% for all authors including the title you may provide
%% the leftmark in one of the following three forms:

%% \renewcommand{\leftmark}
%% {First Author: A Short Title}

%% \renewcommand{\leftmark}
%% {First Author and Second Author: A Short Title}

%% \renewcommand{\leftmark}
%% {First Author et al.: A Short Title}

%% \tableofcontents  % Produces the table of contents.
\section{Background}\label{back}
We consider the problem of estimation and model selection in log-linear
models for large contingency tables involving many categorical
variables. This problem encompasses the estimation of the graphical model
structure for categorical variables. This structure learning task has
lately received considerable attention as it plays an important role in a
broad range of applications. The conditional independence structure of the
distribution can be read off directly from the structure of a graphical
model (a graph) and hence provides a graphical representation of the
distribution that is easy to interpret (see \citet{Lauritzen}). Graphical
models for categorical variables correspond to a class of hierarchical
log-linear interaction models for contingency tables.  Thus, fitting a
graph corresponds to model selection in a hierarchical log-linear model $
\log({\bf p}) = \mathbf{X} \boldsymbol{\beta} $, where $\bf p$ is the
vector of cell probabilities of a table, $\boldsymbol{\beta}$
is a parameter vector and $\mathbf{X}$ is the design matrix (see section
\ref{hl}).

Fitting the log-linear model for large contingency tables in full detail turns out to be a very hard
computational problem in practice. In the following,
we list three possible goals for fitting log-linear models. The goals are
ranked in increasing order according to their computational difficulty:
\begin{description} 
\item [Graphical Structure]Finding the graphical structure for discrete
  categorical variables is easiest but it doesn't allow to infer the
  magnitude of the coefficients $\boldsymbol{\beta}$ in the log-linear
  model, see also formula (\ref{param}).
\item[Parameter Vector $\boldsymbol{\beta}$] The next level of difficulty is the
estimation of the unknown parameter vector $\boldsymbol{\beta}$ in a
log-linear model whose full dimension equals the number of cells in
the contingency table. For large tables, the dimension of
$\boldsymbol{\beta}$ is huge but under some sparsity assumptions it is
possible to accurately estimate such a high-dimensional vector using
suitable regularisation.  The major problem is here that besides the
high-dimensionality of $\boldsymbol{\beta}$, the analogue of the
sample size (the row-dimension of $\mathbf{X}$) is huge, e.g. $3^{40}$
for 40 categorical variables having 3 levels each.  
\item [Probability Vector ${\bf p}$]The most
difficult problem is the estimation of the probability vector ${\bf
  p}$ whose dimension equals again the number of cells in the table.
It is rather unrealistic to place some sparsity assumptions on ${\bf
  p}$ in the sense that many entries would equal exactly zero which
would enable feasible computation.  Therefore, it is impossible to
ever compute an estimate of the whole probability vector ${\bf p}$
(e.g. having dimensionality $3^{40}$).  Nevertheless, thanks to
sparsity of the parameter vector $\boldsymbol{\beta}$ and the junction
tree algorithm, it is possible to compute accurate estimates
$\{\hat{\bf p}(i);\ i \in {\cal C}\}$ for any reasonable-sized
collection ${\cal C}$ of cells in the contingency table. 
\end{description}

There is hardly any method which can achieve all
these goals for contingency tables involving many, say more than 20,
variables. One approach to address the log-linear modeling problem for
large contingency tables is presented in \citet{Jackson}, where some
dimensionality reduction is achieved by reducing the number of levels per
variable. The reduction is accomplished via collapsing two categories by
aggregating their counts if the two categories behave sufficiently
similar. If $d$ variables are considered, this method reduces the problem
at best to $d$ binary variables. For this special case with binary factors,
an approach based on many logistic regressions can be used for fitting
log-linear interaction models whose computational complexity is feasible
even if the number of variables is large \citep{wain}.  Another method to
address the log-linear modeling problem for large contingency tables is
proposed in \citet{Sung}, where the variables are grouped such that they
are highly connected within groups but less between groups and graphical
models are fitted for these subgroups. The subgraph models are then
combined using so-called graphs of prime separators. The implementation of
the combination however is not an easy task and no exact algorithm is given
on how to combine the models.

Our presented methodology allows to achieve all the goals above for
categorical variables having possibly different numbers of levels:
inference of a graphical model for discrete variables, of a sparse
parameter vector in a log-linear model and of a collection of cell
probabilities.  Motivated by the approach in \citet{Sung}, we also propose
a decomposition approach, where the dimensionality reduction is achieved by
recursively collapsing the large contingency table on certain variables
(decomposition) and thereby reducing the problem to smaller tables which
can be handled more easily.  All the fitted lower-dimensional log-linear
models are combined appropriately to represent an estimation of the joint
distribution of all variables.  The procedure enables us to handle very
large tables e.g. up to hundreds of categorical variables, where some or
all of them can have more than two categories. This multi-category
framework is much more challenging than the approach in \citet{wain} for
large binary tables. In \citet{wain2}, an extension to the multi-category
case within the class of pairwise Markov random fields is sketched: it is
argued that higher-order interactions could be included in a pairwise
Markov field but no methodology or algorithm is described how to actually
do this difficult computational task.

\section{Definitions}
In this section, we introduce several important theoretical
concepts. First, we define the general log-linear interaction
model. Subsequently, we introduce the hierarchical log-linear model and the
graphical model as restricted versions of the log-linear interaction
model. Then, we define collapsibility and decomposability, which are
crucial concepts for breaking up a graphical model into smaller pieces.
\subsection{Log-linear interaction model}\label{ll}
We adopt here the notation of \citet{Darroch}. Assume we have some
factors or categorical variables, indexed by a set $V$. Each factor $v \in
V$ has a set of possible levels
$I_v= \{0,1, \ldots,k_v\}$. The contingency table is the cartesian product
of the individual sets: $I=\prod_{v \in V}I_{v}$. An individual
cell in the contingency table is denoted by $i \in I$ and the corresponding cell
count by $n_i$. A marginal index for variable set $a \subseteq V$ is
denoted by $i_a$. For example, assume we have two binary variables, then
$V=\{1,2\}$, $I_1=I_2=\{0,1\}$ and the contingency table is given by
$I=\{(0,0),(0,1),(1,0),(1,1)\}$. An individual cell is for example
$i=(0,1)$ and the index describing the margin along the second variable
($a=2$) is $i_a=i_2=(1)$. The total number of cells in a contingency
table is $m=|I|=\prod_{v\in V}|I_v|$. In our example, $m=4$. A
natural way of representing the distribution of the cell counts is via
a vector of probabilities $\mathbf{p}=(p(i),i\in I)$. In our example,
this would correspond to defining four probabilities $\mathbf{p}=(p_{(0,0)},p_{(0,1)},p_{(1,0)},p_{(1,1)})$. If a total number of $n$ individuals are classified independently, then the distribution of the
corresponding cell counts $\mathbf{n}=(n_1,n_2,\ldots,n_m)$ is
multinomial with probability $\mathbf{p}$. Finally, the general log-linear
interaction model specifies the unknown distribution $\mathbf{p}$
as follows:
\begin{equation}\label{log-linear}
\log p(i)=\sum _{a \subseteq V} \xi_a(i_a) \quad \forall i \in I
\end{equation}
where $\xi_a$ are functions of cell $i$ which only depend on the
variables in $a$. These
functions are called \emph{interactions} between the variables in $a$. If
$|a|=1$, $\xi_a$ is called \emph{main effect}, if $|a|=2$ \emph{first order
interaction} and an \emph{interaction of order $k-1$} if $|a|=k$. For
identifiability purposes we impose constraints on the functions,
namely that $k^{th}$-order interaction functions are orthogonal to
interaction functions of lower order.

\subsection{Hierarchical log-linear models}\label{hl}
A hierarchical log-linear model is a log-linear interaction model with the
additional constraint, that a vanishing interaction
forces all interactions of higher order to be zero as well:
\begin{equation*}
\xi_a=0 \Longrightarrow \xi_b=0 \quad \forall \ a \subseteq b \subseteq V
\end{equation*}
Hierarchical models can be specified via the so-called
\emph{generators} or \emph{generating class} $\mathcal{G}$ which is a
set of subsets of $V$ consisting of the maximal interactions which are
present. More precisely, the generating class $\mathcal{G}$ has the
following property:
\begin{equation}
\xi_a=0 \Longleftrightarrow  \textrm{ there is no } q \in \mathcal{G} \textrm{ with } a \subseteq q.
\end{equation}
Consider an example with three binary factors. An example for a
hierarchical log-linear model is the model consisting of all main
effects, an interaction between $1$ and $2$ and an interaction between
$1$ and $3$: this
corresponds to $\mathcal{G}=\{\{1,2\},\{1,3\}\}$. However, the log-linear
interaction model with
main effect $1$ and interaction between $1$ and $2$ (but no main effect
$2$) is not in the class of hierarchical log-linear models. 

If we go back to formula (\ref{log-linear}) and rewrite it in matrix
formulation, we get:
\begin{equation}\label{param}
\log{(\mathbf{p})}=\mathbf{X}\boldsymbol{\beta},
\end{equation}
where $\boldsymbol{\beta}$ is a vector of unknown 
coefficients and $\mathbf{X} \in \mathbb{R}^{m \times m}$ the design
matrix. Each row of $\mathbf{X}$ corresponds to a certain cell and the
columns of $\mathbf{X}$ correspond to the functions $\xi_a(i_a)$. The
number of columns needed to represent the function $\xi_a$ depends on
the number of different states $i_a$ can take on. For example consider
a categorical variable $a$ that can take on 3 levels. Then, $\xi_a$ is called a
main effect (as $|a|=1$) and $X_a$ (the columns of $\mathbf{X}$
corresponding to $a$) is 2-dimensional.  Originally, it would be
3-dimensional but for identifiability purposes, the subspace spanned by
$X_a$ is chosen orthogonal to the already existing columns of lower
order interaction (here orthogonal to the intercept) and we further choose it orthogonal within the
subspace.  By choosing the identifiability constraints this way, the
parameterization of the matrix used in (\ref{param}) is equivalent to
choosing a poly contrast in terms of ANOVA. If we go back to our
example with two binary factors where $m=4$, (\ref{param}) becomes:
\begin{eqnarray}\nonumber\label{beispiel}
&\log{\mathbf{p}}\begin{pmatrix}
(0,0)\\
(0,1)\\
(1,0)\\
(1,1)
\end{pmatrix}=\mathbf{X}\boldsymbol{\beta},\phantom{blasd}& \\
\\\nonumber
&\textrm{ with } \mathbf{X}=\begin{pmatrix}
  1&\phantom{-}1&\phantom{-}1&-1\\
  1&\phantom{-}1&-1&\phantom{-}1\\
  1&-1&\phantom{-}1&-1\\
  1&-1&-1&\phantom{-}1
\end{pmatrix}, \textrm{ and } \boldsymbol{\beta}=(\beta_{\emptyset},\beta_1,\beta_2,\beta_{12})^T,&
\end{eqnarray}
where the first column is the intercept and belongs to $a=\emptyset$,
the second column belongs to $a=1$ and has entry 1 whenever variable
$a=1$ takes on the first level and -1 else and similarly for the third
column. The fourth column belongs to the interaction between variable
1 and 2.  A description of $\mathbf{X}$ for the general case can be
found in \citet{Ich}. In the following we will denote the components of
$\boldsymbol{\beta}$ belonging to $X_a$ with $\beta_a$.

\subsection{Graphical models}\label{gmsep}
We first introduce some terminology. A graph is defined as a pair
$G=(V,E)$, where $V$ is the set of \emph{vertices} or \emph{nodes} and
$E\subseteq V\times V$ is the set of \emph{edges} linking the vertices.
Each node represents a (categorical) random variable.  Here we only
consider undirected graphs which means that $(u,v) \in E$ is equivalent to
$(v,u)\in E$. A \emph{path} from $u$ to $v$ is a sequence of distinct nodes
$v_0=u,v_1,\ldots,v_n=v \textrm{ such that } (v_i,v_{i+1})\in E $ for all
$i\in\{0,1,\ldots,n-1\}$. Given three sets of variables $A,S,B \subseteq
V$, we say that $S$ \emph{separates} $A$ from $B$ in $V$ if all paths from
vertices in $A$ to vertices in $B$ have to pass through $S$.  Consider a
random vector $\mathbf{Z}=\{Z_v,v\in V\}$ with a given distribution.  We
say that the distribution of $\mathbf{Z}$ is \emph{globally Markov} with
respect to a graph $G$ if for any 3 disjoint subsets $A,S,B \subseteq V$
the following property holds:
\begin{equation}
S \textrm{ separates }A \textrm{ from }B \Longrightarrow Z_A \bot Z_B|Z_S,
\end{equation}
where the symbol ``$\bot$'' denotes (conditional) independence. 
This states that we can read off conditional independence relations
directly from the graph if the distribution is globally Markov with
respect to the graph. Graphical models therefore provide a way to
represent conditional (in)dependence relations between variables in
terms of a graph structure. We say that a set of nodes of $G$ forms a
\emph{complete} subgraph of $G$ if every pair in that set is connected by an
edge. A maximal complete subgraph is called a \emph{clique}.

\begin{figure}
\begin{center}
\caption{Examples of graphical models. (a) Full (maximal) model
  corresponding to the example given by formula (\ref{beispiel}). (b) In
  the same example graph with $\beta_{12}=0$. (c) Example of an interaction
graph corresponding to a hierarchical log-linear model which is not
graphical. (d) The separator $S=2$ has index $\nu(S)=2$. (e) The separator
$S=2$ has index $\nu(S)=1$.}\label{graphen}
\includegraphics[scale=0.3]{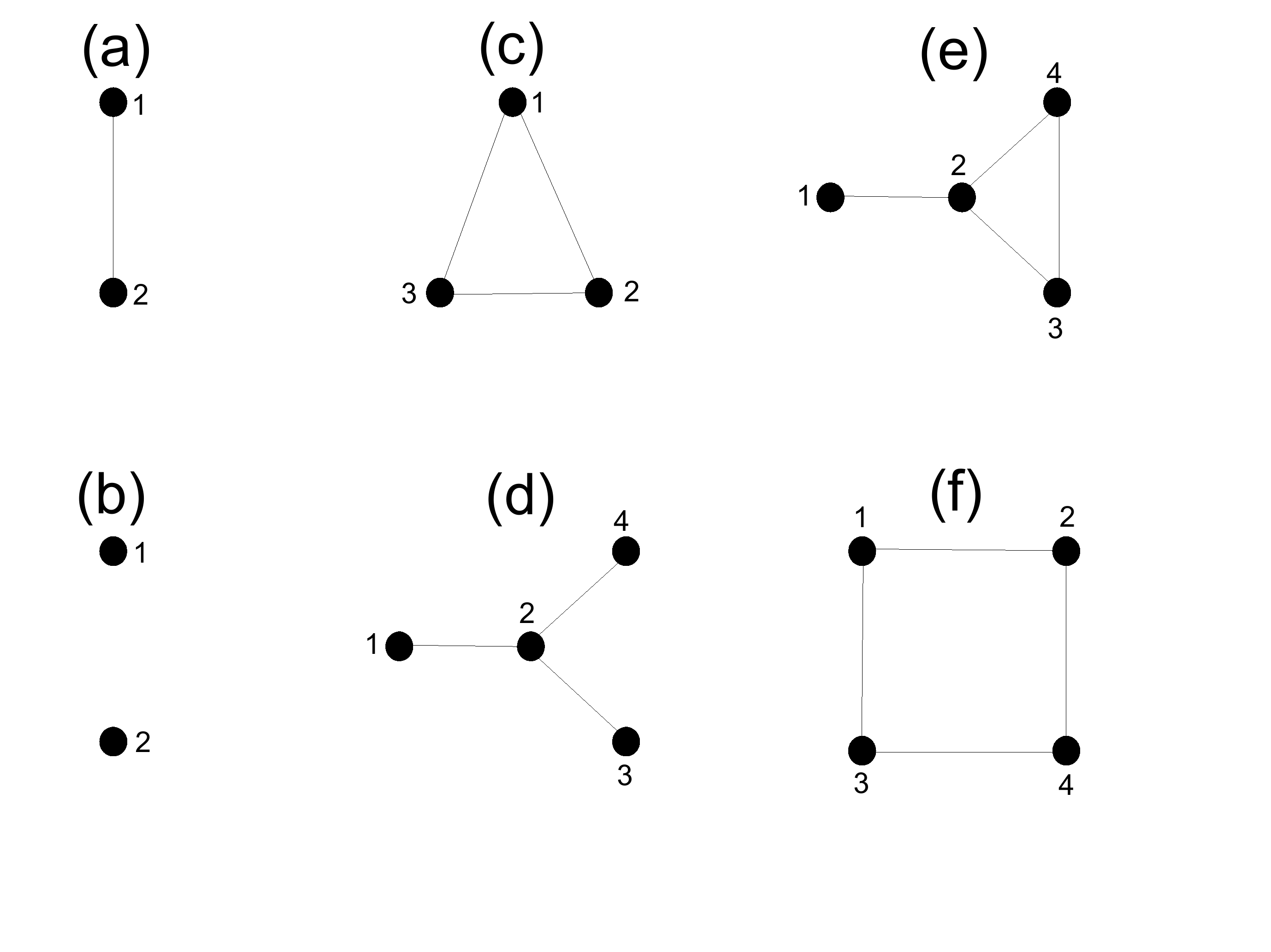}
\end{center}
\end{figure}

The \emph{undirected graphical model} (from now on ``graphical model'' for
short) represented by a graph $G$ corresponds
to a hierarchical log-linear model where the cliques of the graph are
the generators of the model. If we go back to our example in the
previous section and assume that $\beta_{12}\neq 0$ in formula
(\ref{beispiel}), then the hierarchical log-linear model
(\ref{beispiel}) can be represented by the graphical model in Fig. \ref{graphen}(a); if $\beta_{12}=0$, then the
corresponding graphical model is the one in Fig. \ref{graphen}(b).

%Darroch Paper
Conversely, assume that the generators of a log-linear model
are given by a set $\mathcal{G}$. By connecting all the vertices
appearing in the same generator with each other and placing no other edge, the so-called \emph{interaction graph} is built. By the definition of the interaction graph and by looking
at formula (\ref{log-linear}), it becomes clear that the distribution
induced by the log-linear model is Markov with respect to the
interaction graph and we can read off conditional independencies
directly from the graph. It is also clear that $\mathcal{G}$
corresponds to a graphical model via its interaction graph if and only
if $\mathcal{G}$ is the set of cliques of this graph. In that case we
say that $\mathcal{G}$ is a \emph{graphical generating class}. If there are
cliques in the interaction graph which are not in $\mathcal{G}$, the
hierarchical log-linear interaction model is not graphical and its
interaction structure cannot be adequately represented by the graph
alone. However, the graph may still completely represent all
conditional independencies of the underlying distribution. The
simplest example of a hierarchical log-linear model $M$ which is not
graphical is $V=\{1,2,3\}$ and
$\mathcal{G}=\{\{1,2\},\{2,3\},\{3,1\}\}$. Its interaction graph is
shown in Fig. \ref{graphen}(c) which has as its only clique the
complete graph $\{1,2,3\}$. Since the set of cliques and the set of
generators $\mathcal{G}$ are not identical, model $M$ is not graphical.

Any joint probability distribution of discrete random variables can be
expanded in terms of a log-linear interaction model.  For some
distributions it is possible to represent all (conditional)
independencies in an undirected graphical model and these distributions are called
faithful to their interaction graph $G$ or we say that the graph is a
perfect map of the distribution. In other words, the graph captures
all and only the conditional independence relations of the
distribution.

\subsection{Collapsibility}
\emph{Collapsing} over a variable simply means summing over that variable and
thereby reducing (collapsing) the table to the remaining dimensions.

When collapsing over a variable, spurious associations between the remaining
variables may be 
introduced and original associations can vanish. A criterion that addresses
this problem is \emph{collapsibility}. We say that a variable is
collapsible with respect to a specific interaction $\xi_a$, when the
interaction in the original contingency table is identical to the
interaction in the collapsed contingency table (i.e., $\xi_a$ in equation
(\ref{log-linear}) remains the same).\\
The general result regarding collapsibility, which goes back to a
theorem stated in \citet{Bishop}, can be summarized as
follows:
\begin{eqnarray}\nonumber \label{collapsibility}
&& \textrm{\hspace{-1.35cm} By collapsing a table over a variable which interacts
with } s \textrm{ other variables, then  } s \textrm{- and higher} \\\vspace{-0.4cm}
&& \textrm{\hspace{-1.35cm} order interactions between the remaining
variables are not changed in the collapsed
table.}\\\nonumber \vspace{-0.4cm}
&& \textrm{\hspace{-1.35cm} Conversely, lower-order interactions between the
remaining variables are affected by collapsing.}\vspace{-1.4cm}
\end{eqnarray}
\noindent 
Suppose, for example, that variable $X$ only interacts with one other
variable $Y$. If we collapse over $X$, no interaction changes but main
effects may be changed. Furthermore, suppose that $Z$ is independent of all
other variables. Then neither main effects nor interactions
change when collapsing over $Z$.

\subsection{Decomposability}
\label{decomp}
Assume a graphical model on the vertex set $V$. A triple of disjoint subsets
$(A,S,B)$ of the vertex set $V$ forms a \emph{decomposition} if (i)
$V=A\cup S \cup B$, (ii) $S$ separates $A$ from $B$ and (iii) $S$ is complete.

Decomposability is defined recursively: A graph is \emph{decomposable} if it
is complete or if there exists a decomposition $(A,S,B)$ where the
subgraphs of $G$ restricted to the
vertex sets $A\cup S$ and $S \cup B$ are decomposable.

Denote by $\mathcal{C}$ the set of all cliques of a decomposable graph
and by $\mathcal{S}$ the set of all separators. For a decomposable graph with
decomposition into cliques $\mathcal{C}$ and separators $\mathcal{S}$,
the probability of a cell $i$ is given by the following formula (see for
example Proposition 4.18 of \citet{Lauritzen}):
\begin{equation} \label{decomp.formula}
p(i)=\frac{\prod_{C\in \mathcal{C}}p(i_C)}{\prod_{S\in \mathcal{S}}p(i_S)^{\nu(S)}},
\end{equation}
where $\nu(S)$ is the so-called index of the separator. The formal
definition of the index $\nu(S)$ is a bit cumbersome and is given in
\citet{Lauritzen}. However, intuitively it can be thought of as the number
of times the set $S$ acts as a separator. For example: In
Fig. \ref{graphen}(d), node $2$ separates $\{1\}$ and $\{4\}$ (the cliques
consisting of single nodes $\{1\}$ and $\{4\}$), $\{1\}$ and $\{3\}$, and
also $\{3\}$ and $\{4\}$. Therefore the index of the separating node $2$ is
3, as node 2 acts three times as separator. If we look at Fig.
\ref{graphen}(e), we see that the node $2$ only separates $\{1\}$ from the
clique $\{3,4\}$ as the single nodes $3$ and $4$ are no longer cliques
since there is an edge between them. Therefore, the node $2$ only acts once
as separator and thus the corresponding index is 1.

It might not be possible to decompose the graph into decomposable 
components. By definition, this is the case for non-decomposable
graphs. The simplest example of a non-decomposable graph is given in
Fig. \ref{graphen}(f). For a non-decomposable graph one can always 
add a minimal number of edges
to the graph, such that it becomes decomposable (this step is called minimal
triangulation; see \citet{Kristian}). Formula (\ref{decomp.formula}) also holds
for such triangulated graphs.

\section{Estimating a Log-Linear Model Using a Decomposition
  Approach} \label{sec:estimating} In this section, we propose our novel
method for recovering both the graph structure and the parameters of a discrete
graphical model. The underlying idea is to learn models over small sets of
variables and stitch them together. More precisely, we first estimate an
initial graph using node-wise regressions, add triangulations if necessary
to make this graph decomposable and then use the decomposed components as
the smaller sets of variables. The smaller sets are then analyzed one at a
time and stitched together using formula (\ref{decomp.formula}). In the
following, we will describe each step of our method in more detail. The
details of log-linear model selection on the smaller sets of variables are
deferred to section \ref{modelselect}. An outline of our decomposition
approach for log-linear model estimation is given in the display Algorithm
\ref{algo}.

\incmargin{1em}
\restylealgo{boxed}\linesnumbered
\begin{algorithm}\label{algo}
\dontprintsemicolon
\SetKwData{Left}{left}
\SetKwData{This}{this}
\SetKwData{Up}{up}
\SetKwFunction{Union}{Union}
\SetKwFunction{FindCompress}{FindCompress}
\SetKwInOut{Input}{input}
\SetKwInOut{Output}{output}
\caption{Outline of our decomposition approach for log-linear model estimation.}
\Input{Node set $V = \{Z_1,\ldots,Z_p\}$, data matrix $D$ on $V$,
  $\mathcal{C}=\{\}$, $\mathcal{S}=\{\}$, $\mathcal{M}=\{\}$, $s_{max}$}
\Output{Estimated log-linear model}
\BlankLine
\tcp{Estimate importance matrix using node-wise regression (see section
  \ref{sec:InitialGraphEst})}\;
\textbf{Set} $\mathbf{M}$ = $\mathbf{R}$ = $\mathbf{\tilde{R}}$ := $p*p$
matrix consisting of 0's \;
\For{$i$ in 1:p}{
  Do regression $Z_i \sim V \setminus \{Z_i\}$ using \texttt{cforest} on $D$\;
  $\mathbf{M}_{i,-i}$ := importance measure of regression for each covariate \;
  $\mathbf{R}_{i,1:p}$ := ranks of $\mathbf{M}_{i,1:p}$ (small number
  corresponds to small rank)\;
}
$\mathbf{\tilde{R}}_{i,j}$ :=
$\max{(\mathbf{R}_{i,j},\mathbf{R}_{j,i})}$ \;
\BlankLine
\tcp{Triangulation and Recursive Decomposition (see section
  \ref{Decomposition})}\;
$G$ := complete graph on $V$\;
\While{$|V|>0$}{
  \uIf{$G$ is not decomposable}{
    $\tilde{G}$ := minimal triangulation of $G$\;
  }
  \Else{
    $\tilde{G}$ := $G$\;
  }
  find any minimal clique $C$ of $\tilde{G}$\;
  \uIf{$|C| \leq s_{max}$}{
    split $C$ into minimal separator $S$ and rest $A$: $A \cup S = C$\;
    record $C$ in $\mathcal{C}$, $S$ in $\mathcal{S}$\;
    estimate log-linear model on $C$ (see section \ref{decompApproach}) and
    save in $\mathcal{M}$ \; 
    $V$ := $V \setminus A$, $G$ := subgraph of $\tilde{G}$ on $V$,
    $\mathbf{\tilde{R}}$:=$\mathbf{\tilde{R}}_{i \in V, j \in V}$\;
  }
  \Else{
    $(i,j)$ := index of minimal entry in $\mathbf{\tilde{R}}$\;
    $G$ := $G$ where edge between i and j is deleted\;
    $\mathbf{\tilde{R}}_{i,j}$ = $\mathbf{\tilde{R}}_{j,i}$ := $\infty$\;
  }
}
\BlankLine
\tcp{Combination of results (see section \ref{Combination})}\;
$\log{p(i)}=\sum_{C\in \mathcal{C}}\log{p(i_C)}-\sum_{S\in
  \mathcal{S}}{\nu(S)} \log{p(i_S)}$ ($p(i_S)$ and $p(i_C)$ were recorded in $\mathcal{M}$)\;
\end{algorithm}
\decmargin{1em}

\subsection{Estimation of initial graph by node-wise
  regression} \label{sec:InitialGraphEst} If we knew the underlying true
graph and if it was sparse, we could use a decomposition and collapse the
contingency table on sub-tables given by the cliques $\mathcal{C}$ and the
separators $\mathcal{S}$ from the decomposition.  Then we could perform
model selection in the collapsed tables and combine the estimates according
to formula (\ref{decomp.formula}). Of course, we do not know the graph and
therefore we don't know $\mathcal{C}$ and $\mathcal{S}$ for the
decomposition. In this section we propose a method of how to come up with
an initial graph estimate.

A log-linear model measures the associations among the variables. The
association between two variables can also be measured by doing regression
from one variable upon the others. It is thus reasonable to apply a
regression method to find groups of variables which are highly
associated within a group but only weakly associated between groups. 

See for the following also Algorithm \ref{algo} lines 1 to 7. Assume a graphical model on the node set $V=\{1,\ldots,p\}$ with
corresponding random variables $Z=\{Z_1,\ldots,Z_p\}$. For every
node $i$ in the graph, we run a regression with $Z_i$ as response
variable and all remaining variables $V \setminus Z_i$ as covariates. We then draw an edge between
nodes $i$ and $j$ if and only if the covariate $Z_j$ has an influence on
the response $Z_i$. Ideally, the regression method involves
interactions among the covariates (unless we assume a binary Ising model as
in \citet{wain,wain2}). 

Inspired by
\citet{Sung}, we use a non-parametric regression approach. But instead of their single regression tree strategy we use a
Random Forest approach (see \citet{Breiman1}). Trees can naturally incorporate
interactions between variables without running severely into the curse of
dimensionality and are therefore ideal for our purposes. We prefer Random
Forest instead of a single tree, since Random Forest oftentimes yield much more stable results for variable selection.

There are three common ways of measuring the importance of
individual variables in Random Forest regression. First, the importance
measure can be the number of 
times a variable has been chosen as split variable (selection
frequency). Second, the decrease in the so-called Gini index can be
used. Third, the
permutation accuracy which measures the prediction accuracy before and
after permuting a variable can be used.

By performing node-wise regression from each variable on
all others and using any importance measure mentioned above, an importance
matrix $\mathbf{M}$ can be built whose element $M_{i,j}$
describes the importance of variable $Z_j$ (acting as covariate) to the
variable $Z_i$ (acting as response). Note that this matrix is not
symmetric.

%%- Algorithmically: 
%%- \vspace{0.2cm}
%%- \noindent
%%- \textbf{Set} $impat = d \times d $ matrix consisting of 0\\
%%- \textbf{For i in 1:d}\\
%%- \hspace{0.3cm} Do regression $Z_i \sim \mathbf{Z}\backslash Z_i$\\
%%- \hspace{0.3cm} Set $sec:InitialGraphEst[i,-i] <$\hspace{-0.1cm}$-$ importance of regression\\
%%- \textbf{Return} sec:InitialGraphEst 
%%- \vspace{0.2cm}

There is one technical difficulty, which we would like to mention here. A
high entry in the importance matrix indicates a strong association between
the corresponding row and column variable. However, depending on the
importance measure, the values between various predictor variables as well
as between different regressions might not be comparable. It has been shown
that popular importance criteria in Random Forest such as the Gini index,
the selection frequency or the permutation accuracy are all strongly biased
towards variables with more categories (see \citet{Strobl}). We therefore
use the ``cforest'' method proposed by \citet{Strobl} which provides a
variable importance measure that can be reliably used for variable
selection even in situations where the predictor variables vary in their
scale of measurement or their number of categories.

Still, however, the importance measures are only consistent within
rows but not between rows as the variable importance not only depends on
the predictor variables in a regression but also on the response
variable. Therefore, values cannot be directly compared between rows. For
that reason we only consider the ranks of the importance matrix entries
within rows. This yields the importance matrix of ranks $\mathbf{R}$ with
$\mathbf{R}_{i,1:p} = \mathrm{rank}(\mathbf{M}_{i,1},\ldots
,\mathbf{M}_{i,p})$, where small numbers correspond to small ranks.
Thus, two variables $Z_i$ and $Z_j$ are
strongly (conditionally) dependent if and only if $\mathbf{R}_{i,j}$ and
$\mathbf{R}_{j,i}$ are both large. In this context, we define a
symmetrized importance matrix of ranks $\mathbf{\tilde{R}}$ with
$\mathbf{\tilde{R}}_{i,j} = \max (\mathbf{R}_{i,j},\mathbf{R}_{j,i}) \quad \forall i,j \in V$.
An initial graph estimate could now be obtained by placing only those
edges, whose corresponding entry in $\mathbf{\tilde{R}}$ is larger than a
given cutoff.

\subsection{Triangulation and recursive decomposition}\label{Decomposition}
Since it is not clear how to find a suitable cutoff for the initial graph,
we employ a recursive scheme (see Algorithm \ref{algo} lines 8 to
26). First, we start with a complete graph $G_0$. Then, we delete the edge
with the smallest importance according to $\mathbf{\tilde{R}}$. If the
resulting graph $G_1$ is not decomposable, we extend it to its minimal
triangulation $\tilde{G_1}$, which is guaranteed to be decomposable (see
section \ref{decomp}). If $\tilde{G_1}$ has a clique that is small enough
to be analyzed on its own (this depends on computational resources and
could be the case for $s_{max}=10$ binary variables), we split it off. If
it has no such clique, we keep deleting edges in $G_1$ according to
smallest importance (and set the corresponding entry of deleted edges in
$\mathbf{\tilde{R}}$ to infinity, so that it is not considered for deletion
again) until there is a small enough clique. Let's assume that $A \cup S$
corresponds to a clique in the (triangulated) thinned graph, $S$ separates
$A$ from $V \setminus \{A \cup S\}$ and $|A \cup S| \leq s_{max} = 10$. We
then split off this clique (and we then estimate a log-linear model on it, see
section \ref{decompApproach}).

From here, we restart the whole procedure with the remaining graph,
i.e. setting $G_0 := V \setminus A$. As corresponding importance matrix, we
take only those rows and columns of $\mathbf{\tilde{R}}$ which correspond
to the nodes in $V \setminus A$. This is repeated until the remaining graph
consists of cliques whose cardinalities are less or equal to
$s_{max}$. Therefore, the amount of edges which we recursively delete
depends on the maximal size of cliques which we allow, denoted by
$s_{\mathrm{max}}$ which is a tuning parameter of our procedure of initial
graph estimation. As mentioned above, $s_{\mathrm{max}}$ is usually chosen
by computational requirements as the optimal size of the initial graph
seems to be of minor importance.

Note that it is not crucial that we have the sparsest possible subgraphs to
collapse on, as long as any reasonable log-linear model selection procedure
can be applied for these subgraphs. Such a decomposition is implemented in
the R package \texttt{LLdecomp}, available at \texttt{http://www.r-project.org}.

\subsection{Combination of results}\label{Combination}
\begin{figure}\caption{Illustration of how many separator edges to
    take up into the model. x-axis: the fraction of separator interactions
    $\xi$ with $\hat{\beta}_{\xi}=0$ among all separator interactions.
    y-axis: the fraction of non-separator interactions with estimated
    interaction coefficient equal zero.  The points correspond to different
    levels of thresholding. We see that if we threshold 30\% of the
    non-zero $\hat{\boldsymbol{\beta}}$ coefficients, we have almost
    exclusively thresholded separator interactions, as we would
    expect.}\label{howmany}
\begin{center}
\includegraphics[scale=0.4]{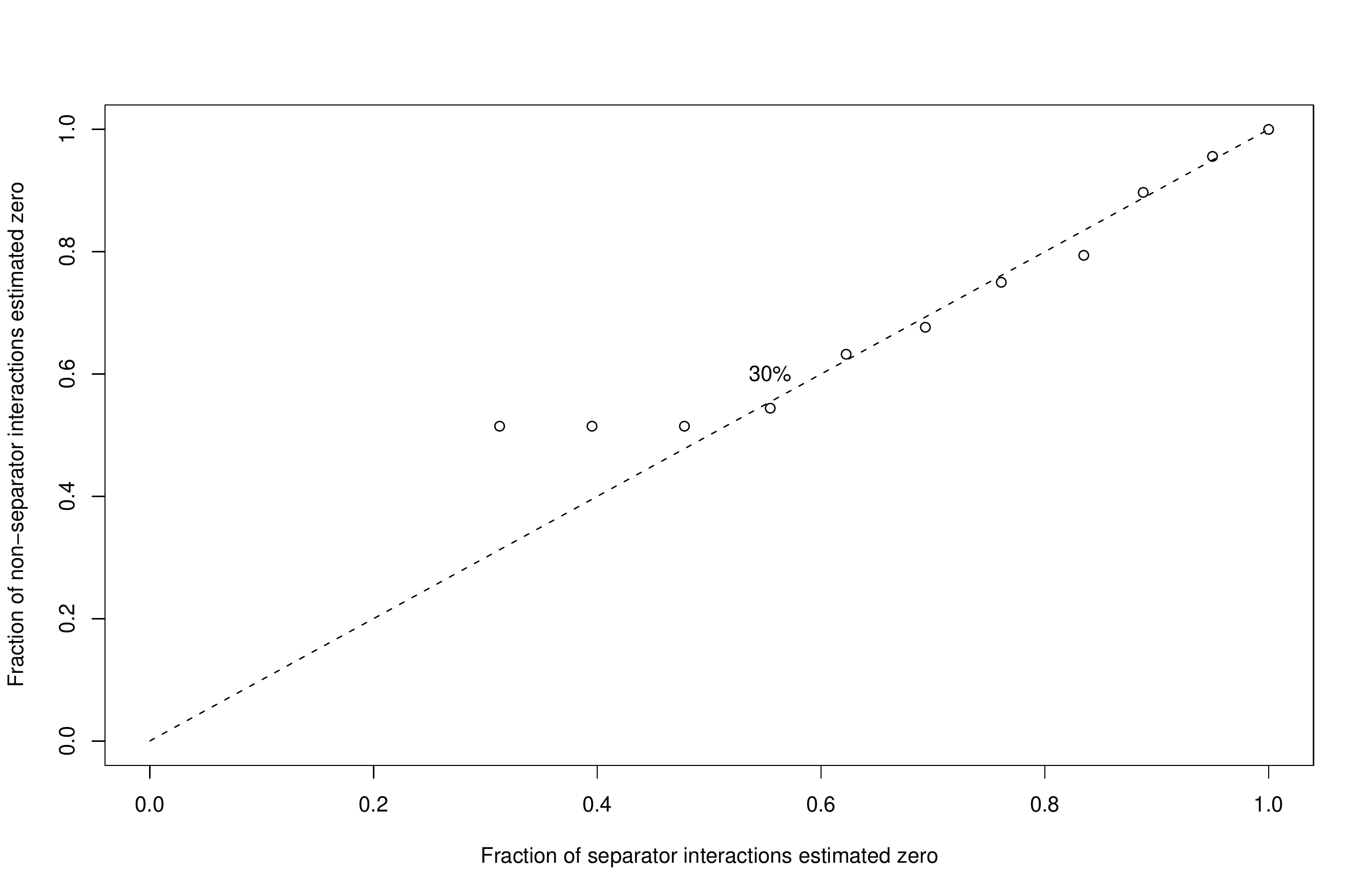} 
\end{center}
\end{figure}

Assume we have collapsed the table on the cliques $\mathcal{C}$ and
separators $\mathcal{S}$ of
the graph induced by the recursive thinning and decomposition procedure. Furthermore,
assume that we have fitted a model for each of these sub-tables (the
collapsed tables on $\mathcal{C}$ and $\mathcal{S}$). We then get the
log-linear model corresponding to the full graph by using formula
(\ref{decomp.formula}):
\begin{eqnarray}
\log{p(i)}&=&\sum_{C\in \mathcal{C}}\log{p(i_C)}-\sum_{S\in
  \mathcal{S}}{\nu(S)} \log{p(i_S)}\nonumber\\
&=& \sum_{C\in \mathcal{C}}X_c\beta_C-\sum_{S\in
  \mathcal{S}}{\nu(S)} X_S\beta_S,\label{combination}
\end{eqnarray}
where $X_C$ and $X_S$ are the design matrices resulting from restricting
the total design matrix to nodes in $C$ and $S$ respectively. The same
notation applies to $\beta_C$ and $\beta_S$.  Formula (\ref{combination})
describes how to aggregate the results of the collapsed tables. In
addition, one can derive from (\ref{combination}) that if we have 3
disjoint subsets $A,S,B$ where $S$ separates $A$ from $B$, then we can
safely collapse over $B$ without changing an interaction between variables
in $A$ or between variables consisting of a mix of $A$ and $S$. The only
interactions which might change are the ones between variables which are
exclusively in $S$ (in the following denoted by separator
interactions). This is in accordance with the result stated in
(\ref{collapsibility}). But as formula (\ref{combination}) holds, the
introduced interactions have a very small $\boldsymbol{\beta}$
coefficient. We therefore expect that if we threshold the parameter vector
$\boldsymbol{\beta}$, most of the introduced zeros belong to so-called
separator interactions $\xi$: $\exists S \in \mathcal{S}$ with $\xi \in S$,
i.e. interactions exclusively contained in a separator.  Consequently, we
set the threshold that the introduced zeros belong to equal parts to
separator- and non-separator interactions. See Figure \ref{howmany} for a
graphical illustration of the procedure. In section \ref{cope} we will
argue empirically that such a thresholding rule works
well. 

\section{Graphical Model Selection Procedures}\label{modelselect}
In this section, we state five methods for log-linear model
selection. Three of them (DGL, DSF and DF) can be used for model selection
within our proposed decomposition approach (see line 19 in Algorithm
\ref{algo}). The remaining two (WW and RF) are established alternative
methods and will be used for comparison.

Our methods use up to three tuning parameters: For decomposition of the
graph, we use a bound on the size of cliques in our graph
($s_{\mathrm{max}}$; see line 16 of algorithm \ref{algo}). Furthermore, in
section \ref{Combination} we introduce a threshold for correcting the
interaction-adding effect of collapsing to marginal models. Finally, in
this section we use regularization parameters for selecting the graph
within components.

\subsection{Model selection for decomposition approach} \label{decompApproach} The first
method we propose is inspired by the Lasso, originally formulated by
\citet{Tibshirani} for estimation and variable selection in linear
regression. Extending this idea, a model selection approach for log-linear
models has been developed in \citet{Ich}. The coefficient vector
$\boldsymbol{\beta}$ is estimated with the group-lasso-penalty
\citep{MingYuan}:
\begin{equation}\label{argmin2}
 \widehat{\boldsymbol{\beta}}^{\lambda}=\textrm{arg}\min_{\boldsymbol{\beta}}\left[-\frac{1}{n}l(\boldsymbol{\beta})+\lambda \mathop{\sum_{a \subseteq C}}_{a\neq 
\emptyset}\lVert \boldsymbol{\beta}_a \rVert_{\ell_2}\right],
\end{equation}
where $l(\boldsymbol{\beta})=\sum_{i=1}^m n_i
(\mathbf{X}\boldsymbol{\beta})_i=\log{\mathbf{p}_{\boldsymbol{\beta}}[\mathbf{n}]}+c$
and
$\lVert\boldsymbol{\beta}_a\rVert_{\ell_2}^2=\sum_{j}(\boldsymbol{\beta}_a)_j^2$. Therefore
$l(\boldsymbol{\beta})$ is up to an additive constant $c$, which does not
depend on $\boldsymbol{\beta}$, the log-likelihood function. This
minimization has to be calculated under the additional constraint that the
cell probabilities add to 1. The group-lasso-penalty has the property
that the solution of (\ref{argmin2}) is independent of the choice of the
orthogonal subspace of $X_a$ and furthermore, the penalty encourages
sparsity at the interaction level. Thus, the vector
$\hat{\boldsymbol{\beta}}_a$ corresponding to the interaction $\xi_a$ (see
section \ref{ll}) has all components either non-zero or zero. Furthermore,
by using group-lasso-penalty model selection we avoid the sampling zero problem,
which is problematic regarding the existence of the MLE (see e.g.
\citet{Christensen}). The tuning parameter $\lambda$ can be assessed by
cross-validation: we divide the individual counts into a number of equal
parts and in turn leave out one part and use the rest to form a training
contingency table with cell counts $\boldsymbol{n}_{train}$.

We abbreviate our decomposition approach using group-lasso-penalty for
arbitrary values of $\lambda$ by DGL (``\textbf{D}ecomposition
\textbf{G}roup \textbf{L}asso''). When $\lambda$ is chosen by
cross-validation, we abbreviate the method by DGL:CV. If, in addition to
that, hard-thresholding for the parameter vector $\boldsymbol{\beta}$ is
used in the specific way described in section \ref{Combination}, we
abbreviate this method by DGL:F (``\textbf{F}inal model of suggested
procedure'').

Our second method is a stepwise forward procedure and aims to minimize the
AIC-type criterion $sk-2\log(l)$, where $l$ is the maximized value of the
likelihood function for the corresponding model with $k$ degrees of
freedom; $s=2$ corresponds to the genuine AIC. Here we also vary the
parameter $s$; a large parameter leads to sparser models. We abbreviate
this method for arbitrary values of $s$ by DSF (``\textbf{D}ecomposition
\textbf{S}tepwise \textbf{F}orward'') and by DSF:AIC if $s=2$.

Finally, we propose a third method where no model selection is performed
after decomposition and the MLE on the decomposed model is used. This
corresponds to DSF with $s=0$. We abbreviate this method by DF
(``\textbf{D}ecomposition \textbf{F}ull model'').

\subsection{Alternative approaches for comparison}\label{alternatives}
Our first method for comparison is given in \citet{wain}, where the problem
of estimating the graph structure of binary valued Markov networks is
considered. They propose to estimate the neighborhood of any given node by
performing $\ell_1$-penalized logistic
regressions on the remaining variables using some penalty parameter
$\lambda$.  Assume we have $d$ binary random variables and observations
thereof $z=(z_1,\ldots,z_d)\in \{0,1\}^d$.  Furthermore, we assume that the
data are generated under the so-called Ising model:
\begin{equation}\label{Ising}
\log{p(\mathbf{z})}= \sum_{s,t=1}^d \beta_{st}z_sz_t+\Psi(\mathbf{B}),
\end{equation}
where $\mathbf{B}$ is a symmetric $d\times d$ matrix and $\Psi(\mathbf{B})$ is a
normalizing constant which ensures that the probabilities add up to one. If
we go back to the log-linear interaction model described in section
\ref{ll} with binary variables, i.e.  the cell $i\in \{0,1\}^d$, then by
comparing formula (\ref{log-linear}) to (\ref{Ising}) we see that the Ising
model is a log-linear model whose highest interactions are of order one and
the parameterization is, in terms of ANOVA, with Helmert instead of poly
contrasts.  Therefore, the interaction graph builds up by connecting the
nodes $s$ and $t$ for which $\beta_{st}\neq 0$. \citet{wain} prove that
under certain sparsity assumptions their method correctly identifies the
underlying graph structure. Note that both our method and \cite{wain} use
node-wise regression. The difference is, that we estimate the node
neighborhoods by performing normal regression and hard-thresholding
importance measures derived from regression weights. We emphasize that the
\citet{wain} approach works for binary variables only, while our
decomposition approach explained in section \ref{sec:estimating} works for
general multi-category variables. For arbitrary
values of the tuning parameter $\lambda$ we abbreviate this method by WW. If
the tuning parameter is chosen by cross-validation, we abbreviate the
method by WW:CV. It turns out, that WW:CV sometimes
yields dense models. Therefore, we abbreviate by WW:MIN the
solution for the minimal $\lambda$ for which the
normalisation constant could be computed (due to computational limitations
connected with the junction tree algorithm).

Our second method for comparison is Random Forest. As explained in section
\ref{sec:InitialGraphEst}, Random Forest together with a suitable cutoff on
the rank importance matrix $\tilde{R}$ can be used to derive an initial
graph estimate. This yields a graph but no estimation of the
log-linear interaction model is provided. We abbreviate this method by
RF.

For convenience, in Table \ref{overview} we give a short overview of the
methods that were defined in this section and are going
to be used in the simulation study in the next section.
\begin{table}[h!]\caption{Overview of the methods used in the simulation studies.}
\label{overview}
\begin{center}
\begin{tabular}{|l|l|}
\hline
Abbreviation & Description\\\hline
DGL& Decomposition approach using group-lasso-penalty without fixing
  the penalty parameter.\\
DGL:CV& As DGL but penalty parameter fixed by cross-validation.\\
DGL:F& Final result of DGL and hard-thresholding as explained in
  section \ref{Combination}.\\
DSF& Decomposition approach using stepwise forward selection without
  fixing penalty parameter.\\
DSF:AIC& As DSF but penalty parameter fixed corresponding to AIC.\\
DF& Decomposition approach without model selection but using MLE.\\
WW& Approach by \cite{wain} without fixing the penalty parameter.\\
WW:CV& As WW but penalty parameter fixed by cross-validation.\\
WW:MIN& As WW using minimal penalty parameter that is computationally
  feasible for junction tree.\\
RF& Random Forest.\\
\hline
\end{tabular}
\end{center}
\end{table}

\section{Simulation Study}\label{simstudy}
\begin{figure}\caption{Graph from which we simulate. Nodes correspond to
    binary random variables.}\label{simulatedgraph}
\begin{center}
\includegraphics[scale=0.5]{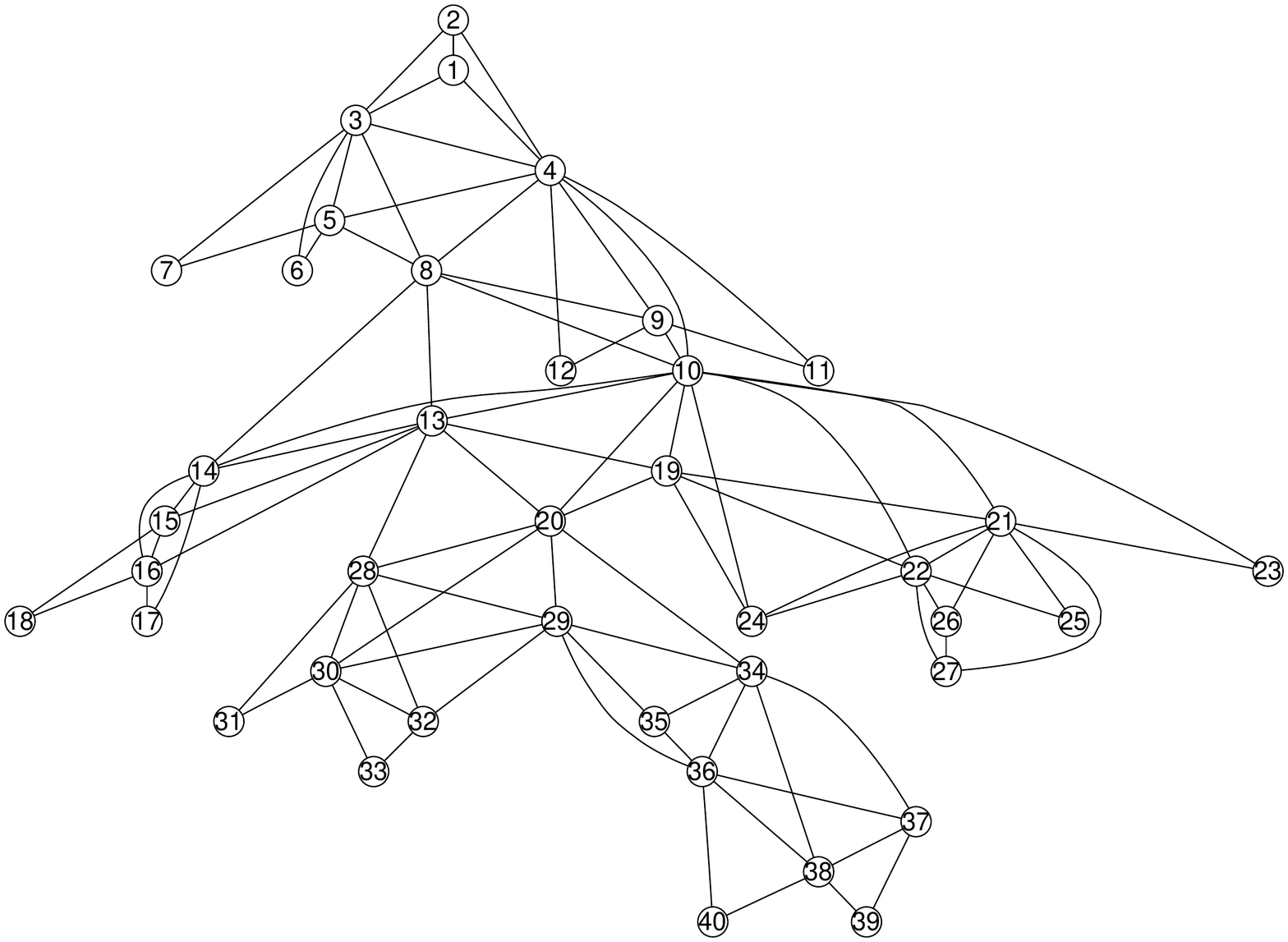}
\end{center}
\end{figure}

We simulate from a log-linear interaction model corresponding to a graph
with 40 nodes and 91 edges. Each node corresponds to a binary variable (and
thus, we can compare with the method in \citet{wain}, explained in section
\ref{alternatives}). The graph is displayed in Fig. \ref{simulatedgraph}. This is
the same simulation setting as was used in \citet{Sung}.
We generate 10 datasets each consisting of 100000 observations
according to the graph in Fig. \ref{simulatedgraph}. 

In section \ref{dcsize}, we investigate the effect of the maximal clique
size $s_{max}$ (for splitting off a clique as explained in section
\ref{Decomposition}) in our decomposition approach with respect to
performance in estimating the correct model structure. In section
\ref{gmsp} we compare different approaches with respect to performance in
estimating the correct model structure, while in section \ref{cope} we
compare in terms of accuracy for estimating the cell probabilities.

\subsection{Optimal clique size} \label{dcsize} 
\begin{figure}[h!]\caption{Comparison of decomposition sizes. Decomposition
    into cliques of maximal size 3, 5 and 10 using DGL:CV. The curves
    corresponds to models which arise by thresholding the final
    $\hat{\boldsymbol{\beta}}$-coefficient.}
\label{decompsize}
\begin{center}
\includegraphics[scale=0.4,angle=-90]{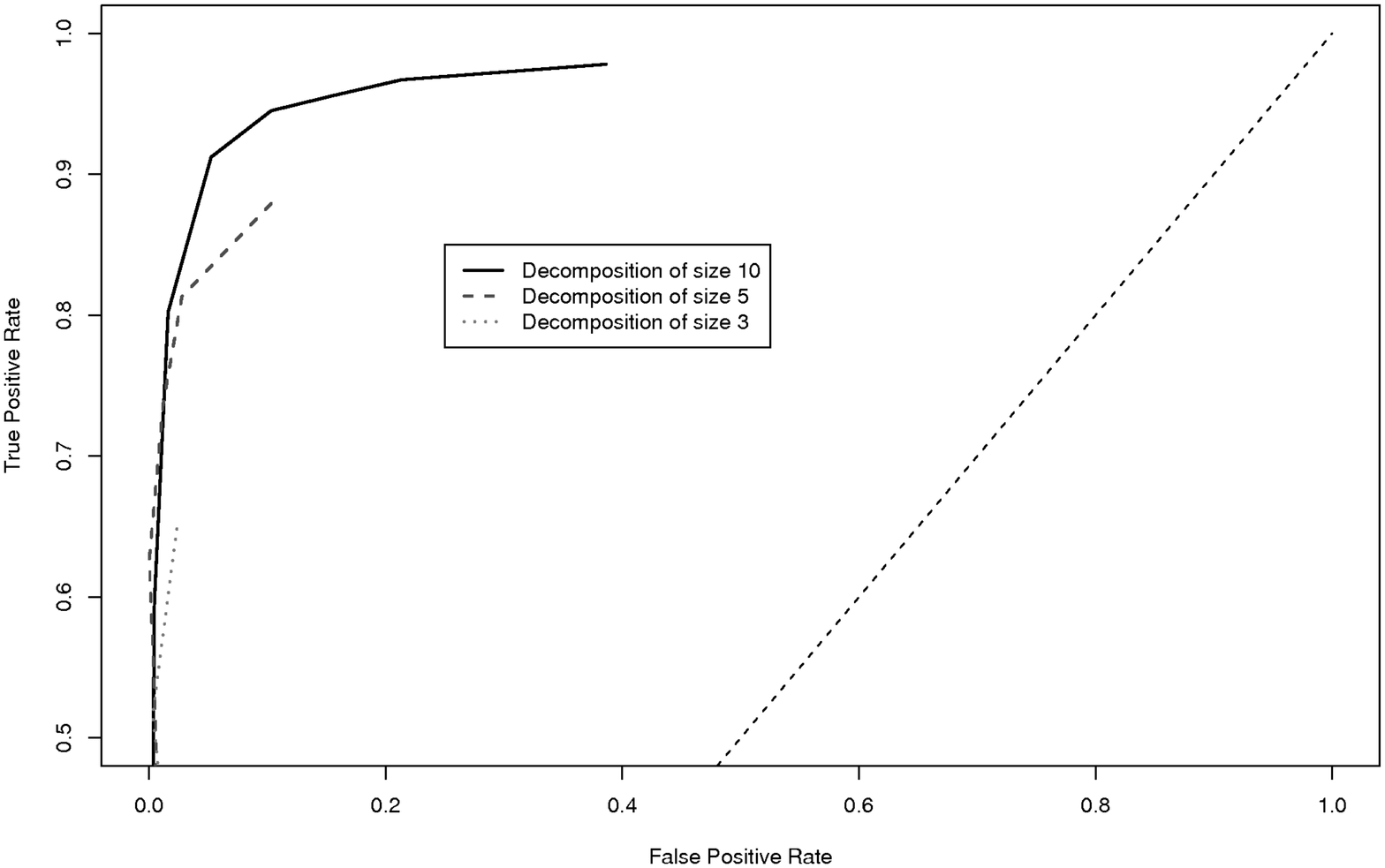}
\end{center}
\end{figure}

We estimate a model using
DGL:CV for $s_{max}$ equal to 3, 5 and 10. A ROC curve is shown in
Fig. \ref{decompsize}, where the endpoints of the curves correspond to the
selected model of DGL:CV. The curves then build up by successively
eliminating edges corresponding to the smallest estimated interaction
vector coefficient $\boldsymbol{\hat{\beta}}$. We see here that larger
decomposition sizes lead to slightly more favorable ROC curves. The picture
remains qualitatively the same if we use DSF instead of DGL. For the
remainder of the simulations, we will keep the maximal clique size in our
approach fixed at $s_{max}=10$.

\subsection{Performance for structure estimation} \label{gmsp} 
\begin{figure}[h]\caption{Comparison of decomposition approaches DGL, DSF
    and DF using $s_{max}=10$. The dotted vertical line corresponds to the
    difference in the true positive rate (number of correctly selected
    edges / number of true edges) for the two procedures when they are
    compared at the false positive rate (number of incorrectly selected
    edges / number of true gaps) of DGL:F. Note that the x-axis is shown
    only up to $0.5$.}
\label{decompare}
\begin{center}
\includegraphics[scale=0.4,angle=-90]{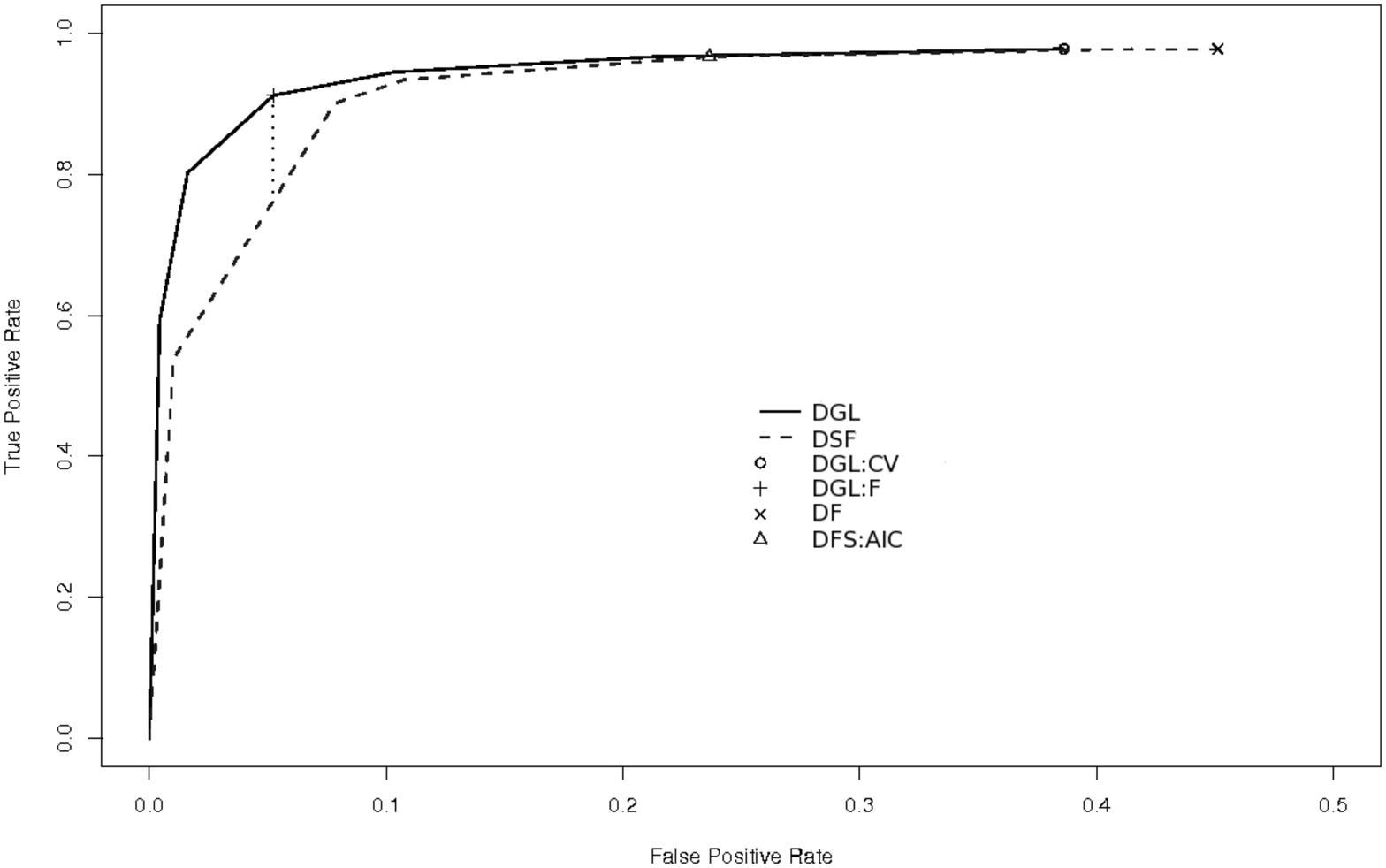}
\end{center}
\end{figure}

First, we compare DGL, DSF and DF. The performance for
structure estimation is shown in ROC curves (see Fig. \ref{decompare}). For
DSF the curve builds up by varying $s$ (compare section
\ref{decompApproach}). Note, that the starting point of the curve of
DSF coincides with DF. DGL starts from DGL:CV and
uses hard-thresholding of $\hat{\boldsymbol{\beta}}$ for obtaining the
values on the ROC curve. Note that during this hard-thresholding, the value
of DGL:F is produced, too.  We see that DSF and
DGL lead to models which have approximately the same number of
false positive and false negative edges, but the DGL is slightly
favorable. The solution of DF has the largest false positive rate (as was
expected since no model selection was done). 

\begin{figure}[h]\caption{Comparison of DGL with WW and RF. For DGL the
    line builds up by thresholding the
    $\hat{\boldsymbol{\beta}}$-coefficient where $\lambda$ is chosen by
    cross-validation. For RF, the edges with least importance are
    successively eliminated and for WW, the tuning parameter $\lambda$ is
    varied. The grey vertical line indicates the false positive rate of
    DGL:F at which the true positive rates of the other methods are compared in
    Table \ref{differenz}.}
\label{globalcompare}
\begin{center}
\includegraphics[scale=0.4,angle=0]{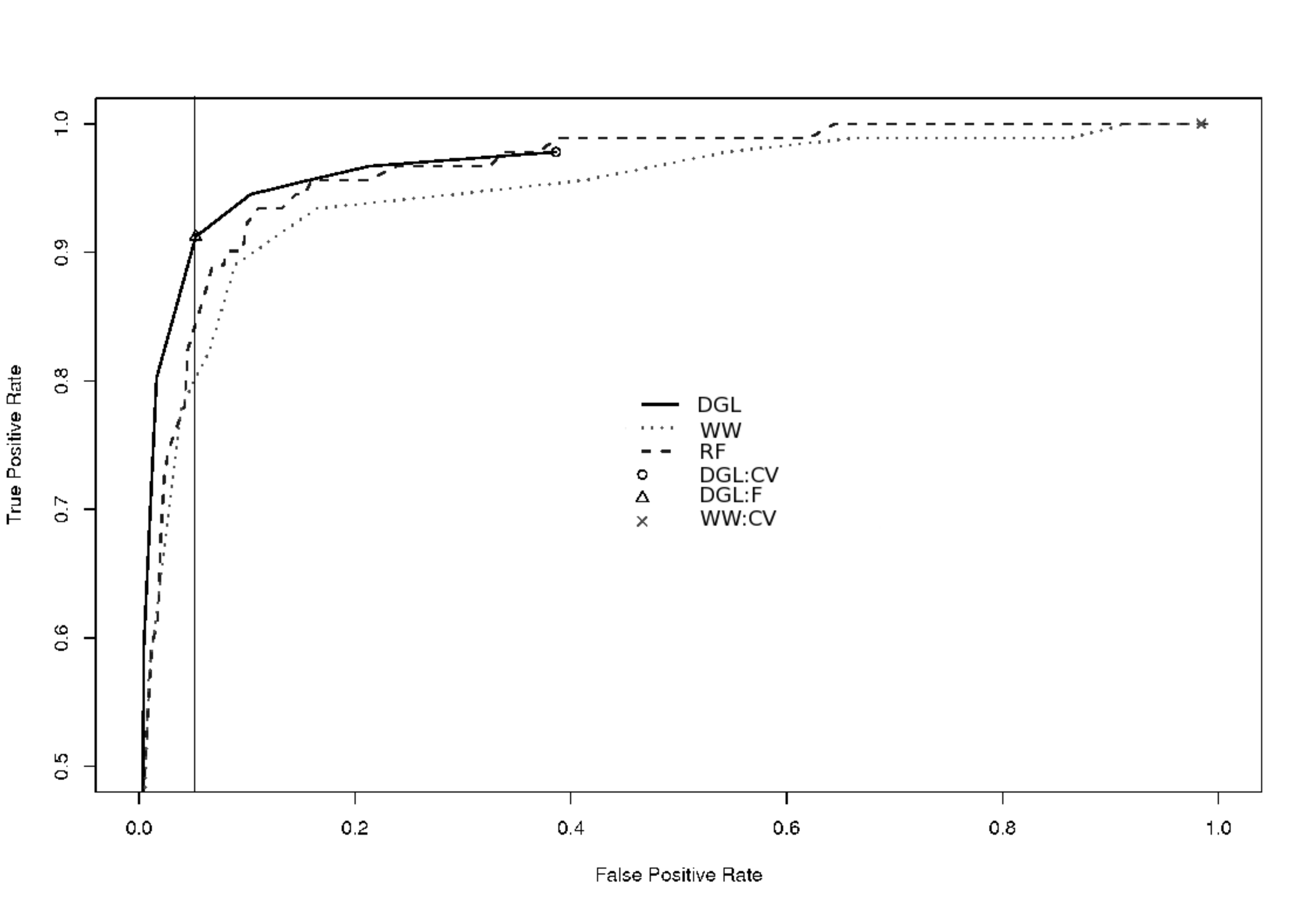}
\end{center}
\end{figure}

Second, in Fig. \ref{globalcompare} the alternative approaches WW and RF
are compared with DGL. In order to keep a simple overview, the results for
DSF are no longer shown. We see that our decomposition approach (DGL)
slightly outperforms the alternative approaches WW and RF. Keep in mind
that in addition to the advantage of our method in performance, RF does not
yield an estimate for the parameter vector and WW does so only for binary
variables, whereas our method is applicable to general multi-category
variables.

Even though Fig. \ref{decompare} and Fig. \ref{globalcompare} only
represent one simulated dataset, we observed a similar picture for other
simulation settings.  The lines for DGL and DSF are always very close, with
DGL being slightly better and both methods being clearly superior to the
global approaches. The reason why Fig.  \ref{decompare} and
Fig. \ref{globalcompare} only display results from one dataset is that the
single final models cannot be averaged over different datasets as they have
different positions on the curves for different datasets (different numbers
of true and false positives). If we average over all these values, the
result is not very meaningful anymore.  However, we can average the
differences of true positive rates for e.g. DGL and DSF where both methods
have the same number of false positives as the solution of DGL:F (dotted
vertical line in Fig.  \ref{decompare}). The results of such comparisons
are summarized in Table \ref{differenz}. We see that DGL yields a
significantly (for significance level $\alpha=0.05$) higher true positive
rate than the corresponding solutions of DSF and WW. On the other hand, the
comparison between the solution of DGL and RF (at the false positive rate
of DGL:F) shows no significant difference.

\begin{table}\caption{Comparison of true positive rates of different
    methods at the false positive rate of DGL:F (indicated in
    Fig. \ref{decompare} and Fig. \ref{globalcompare} by vertical lines).}
\label{differenz}
\begin{center}
\begin{tabular}{|l|c|c|}\hline
&Mean difference&p-value (t-test)\\\hline
tpr(DGL) - tpr(DSF)&0.052&0.036\\\hline
tpr(DGL) - tpr(WW)&0.060&0.011\\\hline
tpr(DGL) - tpr(RF)&0.013&0.256\\\hline
\end{tabular}
\end{center}
\end{table}

\subsection{Performance for estimation of cell probabilities}\label{cope}
DGL, DSF, DF and WW yield estimates of the parameter
vector $\boldsymbol{\beta}$ which can be immediately transformed into
estimates of cell probabilities. In this
section we will compare these methods with respect to the performance for
estimating the cell probabilities. Recall that RF does not yield an
estimate of the parameter vector and is therefore not included in this
comparison.

All approaches considered in this section yield the parameter vector
$\boldsymbol{\beta}$ only up to a constant. We need to ensure that the
estimated cell probabilities add up to one. For sparse graphs, we can use
the junction tree algorithm to calculate the normalising constant for the
probabilities. For a detailed description see \citet{Lauritzen}.

We compare the estimated probabilities, using cross-validation for tuning
parameters and using an expression which is up
to a constant the Kullback-Leibler divergence between the estimated
and the true probability (non-normalized Kullback-Leibler divergence):
\begin{equation}
-\log{\left(\prod_{i}
    \hat{p}_{i}^{p_i}\right)}=-\sum_{i}p_i\log{\hat{p}_{i}},
\end{equation}
where $\mathbf{\hat{p}}$ is the estimated probability vector and
$\mathbf{p}$ denotes the true probability vector. As this sum requires
the calculation of $2^{40}\approx 10^{12}$ components of
$\mathbf{\hat{p}}$ and $\mathbf{p}$ and the summation of the two huge
vectors, this is computationally not feasible. To avoid this problem,
we calculate an empirical version by simulating one million
observations from the graph in Fig. \ref{simulatedgraph} and summing
over these $10^6$ values only.

\begin{figure}[h]\caption{Mean empirical (non-normalized) Kullback-Leibler
    distance between true and estimated probability in dependence of the
    percentage of thresholded coefficients. The vertical line indicates the
    average of percentages of thresholded coefficients using the
    thresholding rule from section \ref{Combination}.}
\label{KL2}
\begin{center}
\includegraphics[scale=0.4,angle=-90]{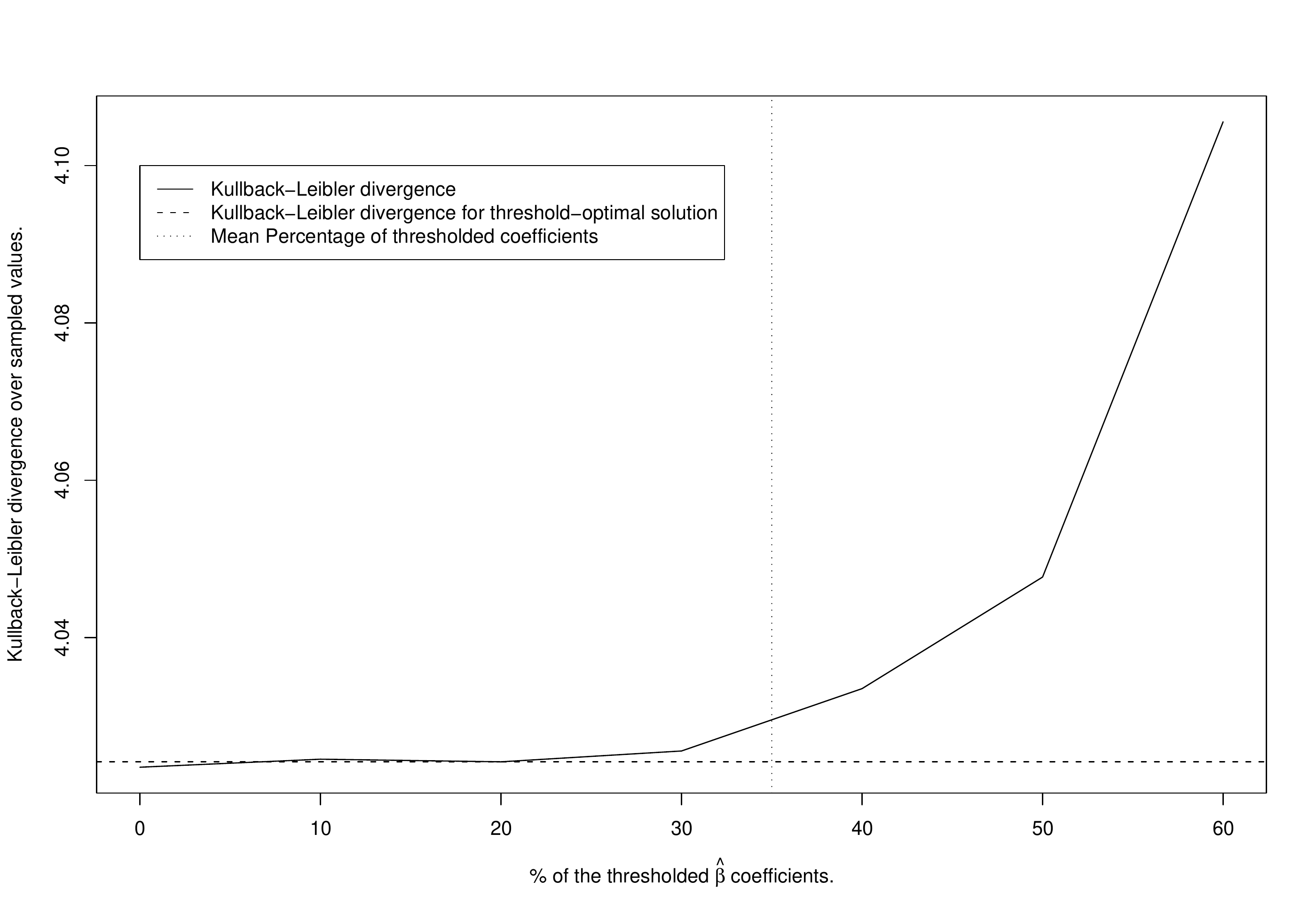}
\end{center}
\end{figure}
The results are summarized in Table \ref{likelihoodcomp}. We see
that DGL:F, DSF:AIC and DF perform similarly and WW:MIN is clearly
inferior. For WW:CV, which is indicated in Fig. \ref{globalcompare}, the
normalizing constant could not be computed. This is because the CV-optimal
solution almost corresponds to the full model and thus, the junction tree
algorithm is not feasible. The maximal computable solutions of WW still
correspond to very large models, which on average involve $22.05\%$ of all
possible edges, compared to $17.01\%$ for DGL:F and $11.66\%$ for the true
graph.

Table \ref{wilcox} provides further insight about significance of the
differences in Table \ref{likelihoodcomp}. All methods are compared against
each other using a paired t-test for the empirical (non-normalized)
Kullback-Leibler divergences and the $p$-values are reported. One can see
that there is no significant (at significance level $\alpha=0.05$)
difference for the decomposition approaches, whereas they are all superior
to WW. This provides evidence that the decomposition of the model is more
crucial than the effective choice of the log-linear model fitting procedure
afterwards.

Furthermore, it is worthwhile stating that the thresholding of the
coefficients (see section \ref{Combination}) does hardly influence the
likelihood as can be seen in Fig. \ref{KL2}. On average, $35\%$ of the
coefficients are thresholded as indicated by the dotted line.  However, for
these threshold-optimal solutions, calculated as described in section
\ref{Combination}, the empirical (non-normalized) Kullback-Leibler
divergence is approximately the same as for the non-thresholded model.

\begin{table}[h!]\caption{Mean empirical (non-normalized) Kullback-Leibler
    divergence between true and estimated probabilities. The $p$-values for
    all the pairwise comparisons are given in Table \ref{wilcox}.}
\label{likelihoodcomp}
\begin{center}
\begin{tabular}{|l|c|c|}
\hline
&Mean&SD\\\hline
DGL:F & 4.0080&0.0204\\\hline
DSF:AIC &4.0242&0.0217\\\hline
DF &4.0223&0.0099\\\hline
WW:MIN &4.3360&0.1094 \\\hline
\end{tabular}
\end{center}
\end{table}
\begin{table}[h!]\caption{All possible pairwise comparisons between models:
    $p$-values of a paired t-test for the equality of (non-normalized)
    Kullback-Leibler divergence.}
\label{wilcox}
\begin{center}
\begin{tabular}{|c|c|c|c|}
\hline
&DSF:AIC&DF&WW:MIN\\\hline
DGL:F&0.1030 &0.0677&4.7 $10^{-6}$\\\hline
DSF:AIC&-&0.8080&6.0 $10^{-6}$ \\\hline
DF&-&-&7.5 $10^{-6}$ \\\hline
\end{tabular}
\end{center}
\end{table}
\section{Application to Tissue Microarray Data}
\subsection{Tissue Microarray technology }
The central motivation that led to this work was to fit a graphical
model to discrete expression levels of biomarkers resulting from
Tissue Microarray (TMA) experiments. Tissue Microarray technology allows
rapid visualization of molecular targets in thousands of tissues at a
time, either at DNA, RNA or protein level. Tissue Microarrays 
are composed of hundreds of tissue sections from different patients
arrayed on a single glass slide. With the use of immunohistochemical
staining, they provide a high-throughput method to analyze potential
biomarkers on large patient samples. The assessment of the expression
level of a biomarker is usually performed by the pathologist on a
categorical scale: expressed/not expressed, or the level of
expression.

Tissue Microarrays are powerful for validation and extension of findings
obtained from genomic surveys such as cDNA microarrays. cDNA microarrays
are useful to analyze a huge number of genes, e.g. a couple of thousands in
one specimen at a time.  In contrast, TMAs are applicable to the analysis
of one target at a time, denoted as biomarker, but in up to 1000 tissues on
each slide. 

The analysis of the interaction pattern of these biomarkers and in
particular the estimation of the graphical model associated with the
underlying discrete random variables are of bio-medical importance.  These
graph-based patterns can deliver valuable insight into the underlying
biology. A detailed description of the Tissue Microarray technology can be
found in \citet{Sauter}.

\subsection{Estimation of graphical model}
Our TMA dataset consists of Tissue Microarray measurements from renal
cell carcinoma patients. We have information from 1116 patients, 831
thereof having a clear cell carcinoma tumor, which is the tumor of interest here.
We have identified 18 biomarkers from which we have information for
the majority of the patients. Among 831 ccRCC (clear cell renal cell
carcinoma) observations, 527 observations are complete with all
biomarker measurements available.  For 87 observations one measurement
was missing, 64 and 30 observations had 2 or 3 missing values,
respectively. 123 observations contained more than 3 missing values
and were ignored in the analysis.  For the observations with 1, 2 or 3
missing values, multiple imputation was applied using the R package
mice \citep{mice}. From 18 biomarkers, 9 are binary and 9 have 3
levels.

\begin{figure}[h!]\caption{Estimated graphical model from Tissue Microarray data}\label{grm}
\begin{center}
\includegraphics[scale=0.5]{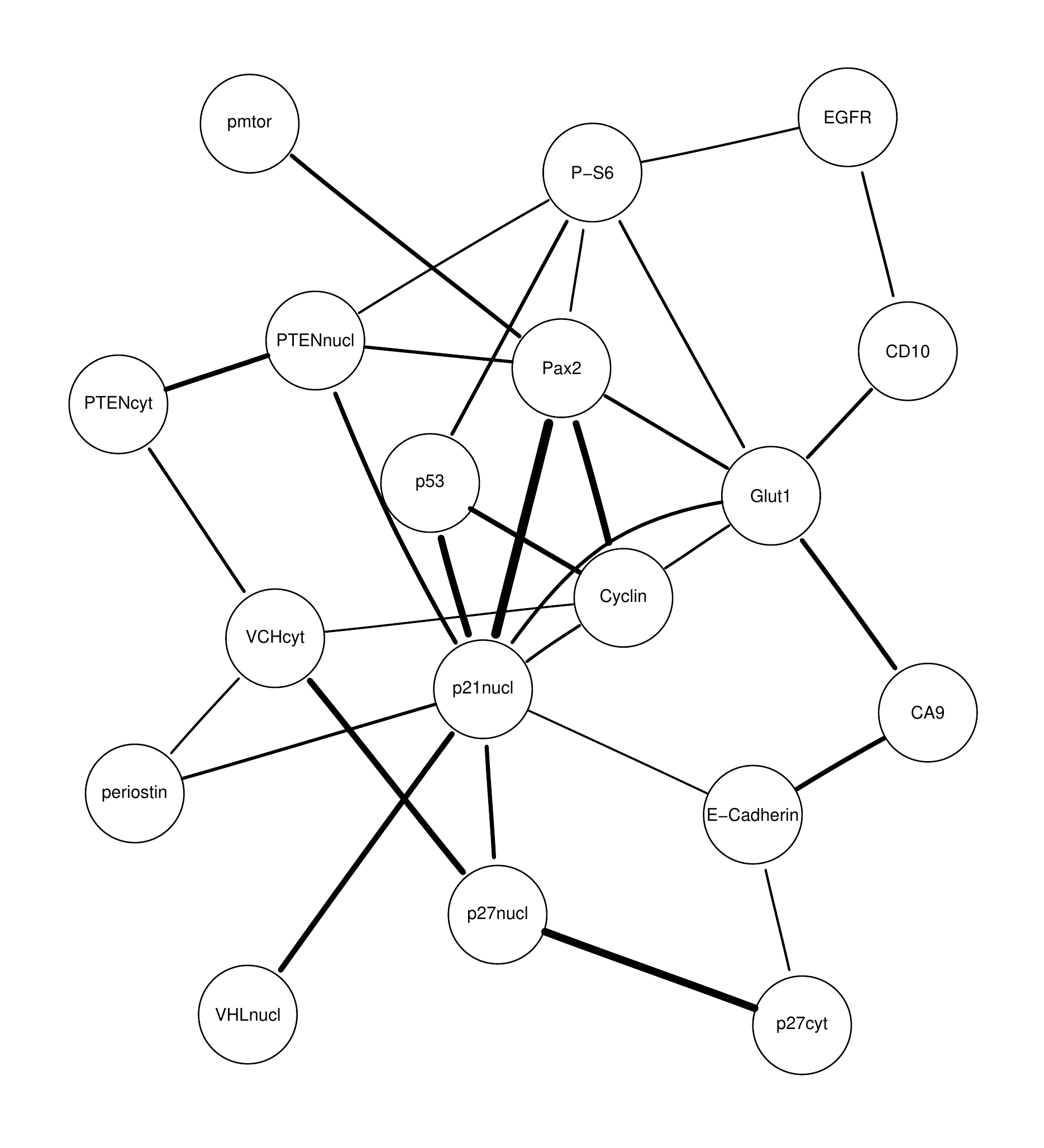}
\end{center}
\end{figure}

Using DGL:F, we estimated a graphical model to the TMA data. The resulting
graph is displayed in Fig. \ref{grm}. The thickness of the line corresponds
to the $\ell_2$-norm of the respective interaction coefficients. Two
biomarkers connected by a thick line, as is the case for nuclear p27 and
cytoplasmic p27, indicates a strong interaction. The kinase inhibitor p27
exhibits its function in the cell nucleus and therefore recent studies have
focused on nuclear p27 expression. Our graphical log-linear model however
shows a tight association between nuclear and cytoplasmic expression of
p27. Therefore it can be speculated that both nuclear and cytoplasmic
presence is required to ensure proper function of p27. It has been shown
that in renal tumors, the von Hippel-Lindau protein (VHL protein) is
upregulating the expression of the tumor suppressor p27 \citep{Osipov}. The
graphical model here provides supporting evidence that VHL indeed regulates
p27, and the corresponding $\boldsymbol{\beta}$ coefficient (not displayed
here) implies that it is a positive regulation.

Furthermore it has been shown in vitro by \citet{Roe} that VHL
increases p53 expression which is a tumor suppressor. In our model it
seems as if p53 is conditionally independent of VHL.  Indeed, it has
long been known that p53 activates expression of p21 (e.g.
\citet{Kim}).  This dependence is displayed very clearly in the
graphical model. We can therefore view the p53-p21 pathway with its
strong interaction as one unit and it is therefore very reasonable
that nuclear VHL interacts with p53. As nuclear VHL is only expressed
in 14\% of the tumors, and it further makes sense from a biological
point of view that the strong interaction between VHL and the p21-p53
pathway is in fact a causal relation, we can indeed speculate that the
loss of VHL deactivates the tumor suppressor p53 which in turn favors
tumor development.
 
CA9, Glut1 and Cyclin D1 are all hypoxia-inducible transcription
factor (HIF) target genes \citep{Wenger}. HIF has not been
measured but we can clearly see that all these HIF targets are
connected by a rather thick line implying that they might react to a
common gene. In addition, CD10 strongly interacts with Glut1 a known
HIF target which suggests that CD10 might also be regulated by HIF.
The reduction of E-Cadherin expression has been found to be negatively
correlated with HIF expression in \citet{tsumo}. This is supported by
a strong negative interaction between E-Cadherin and CA9 which is
positively correlated with HIF expression (not measured).
 
A lot of supporting evidence has been delivered for already existing
theories. However, two strong interactions, one between PAX2 and
nuclear p21 and the other between PAX2 and Cyclin D1 cannot be
immediately explained. PAX2 is absent in normal renal tubular
epithelial cells but expressed in many clear cell renal cell carcinoma
tumours (see \citet{Mazal}). Its frequent expression together with the
strong interaction with the p21-p53 pathway, Cyclin D1 and PTEN make
PAX2 an interesting and possibly important molecular parameter whose
exact function and role still remains to be elucidated.

A more detailed discussion of bio-medical implications is given in \cite{Dahinden09}.

\section{Discussion}
We have proposed a decomposition approach to estimating log-linear models
for large contingency tables and for fitting discrete graphical models. In
a simulation study we have compared various algorithms and concluded that
our procedures are very competitive. It seems that the decomposition of the
problem is much more crucial than the choice of the algorithm to handle the
smaller decomposed datasets: no matter whether DGL, DSF or DF is applied,
the resulting models are superior to non-decomposition approaches such as
WW and RF for model selection as well as for probability or parameter
estimation.

Maybe most important is the computational feasibility of our procedure for
large contingency tables. The
proposed method is scalable to orders of realistic complexity (e.g. dozens
up to hundreds of factors) where most or all other existing algorithms
become infeasible. In particular, our procedure is not only capable of
handling binary data but can easily deal with factors with more
levels. Furthermore, with DGL one doesn't risk the nonexistence of the
parameter estimator in case of sampling zeroes in the contingency table as
this might arise when using the MLE. Our procedure not only fits a
graphical model but also yields an estimation of the parameter vector
$\boldsymbol{\beta}$ in a log-linear model and therefore of the cell
probabilities.  All this is achieved with good performance in comparison to
other methods.  As a drawback, if the true underlying graph has a clique
which is larger than our decomposition size, then some of the edges in the
graph are necessarily lost.

We apply the proposed approach to a problem in molecular biology and we
find supporting evidence for dependencies between biomarkers which have
already been found to exist in vitro or some even in renal tumors, the
domain of our application. However, some strong interactions cannot be
explained immediately and therefore, new biological hypotheses arise.

The R package \texttt{LLdecomp} for computing a decomposition as described
in this paper (i.e., recursively finding cliques and separators) is available at
\texttt{http://www.r-project.org}.

\begin{acknowledgement}
  We would like to thank an Associate Editor and two referees for their
  constructive comments. We would also like to thank Dr. Peter Schraml for
  valuable comments, suggestions and discussions regarding biological
  interpretation.
\end{acknowledgement}

\vspace{0.5cm}

\noindent\textbf{Conflict of Interest Statement}\\
The authors have declared no conflict of interest.

\bibliographystyle{rss}  % Style BST file
\bibliography{refs4}

\end{document}